\documentstyle[11pt,newpasp,twoside,psfig,epsfig]{article}
\def\Mpc{\ifmmode {\, h^{-1} \, {\rm Mpc}}
\else {$h^{-1}\,$ Mpc}\fi}

\def\s8{{\sigma_8}}

\def\ltsima{$\; \buildrel < \over \sim \;$}
\def\simlt{\lower.5ex\hbox{\ltsima}} 
\def\gtsima{$\; \buildrel > \over \sim \;$} 
\def\simgt{\lower.5ex\hbox{\gtsima}} 
\def\bfw{{\bf w}}

\def\omegab{{\Omega_{\rm b}}}

\def\qrms{{Q_{\rm rms}}} 
\def\omegabh2{{\omegab h^2}} 

\def\edcomment#1{\iffalse\marginpar{\raggedright\sl#1\/}\else\relax\fi}
\reversemarginpar

\begin{document}
\title{Cosmological Parameters and Hyper-Parameters: 
The Hubble Constant from Boomerang and Maxima}
\author{Ofer Lahav}
\affil{Institute of Astronomy, Madingley Road,
Cambridge CB3 0HA, UK}

\begin{abstract}
We generalise the procedure for joint estimation of cosmological parameters
to allow freedom in the
relative weights of various probes.  This is done by including in the
joint Likelihood function a set of `Hyper-Parameters', which are dealt
with using Bayesian considerations.  
The resulting algorithm is 
simple to implement.  We illustrate the method by estimating the
Hubble constant $H_{0}$ from the  recent Cosmic Microwave Background
experiments Boomerang and Maxima.
For an assumed flat $\Lambda$-CDM model with fixed
parameters 
$(n=1, \Omega_m = 1 -\lambda = 0.3, \Omega_b h^2 = 0.03, \qrms = 18 \mu K)$
we solve for a single parameter,  
$H_{0}= 79 \pm 4$ km/sec/Mpc
(95 \%  CL, random errors only), slightly higher but still  
consistent with recent results from Cepheids.
We discuss how the `Hyper-Parameters' 
approach can be generalised for a combination of cosmic probes, and
for other priors on the Hyper-Parameters.
\end{abstract}

\section{Introduction}

Several groups  (e.g. Gawiser \& Silk 1998; Webster et al. 1998; 
Lineweaver 1998; 
Eisenstein, Hu \& Tegmark 1999; Efstathiou et al. 1999;  
Bridle et al. 1999, 2000;  Bahcall et al. 1999) 
have recently discussed the estimation
of  cosmological parameters by joint analysis 
of data sets, e.g. Cosmic Microwave Background (CMB), 
SNe Ia, redshift surveys, cluster abundance
and peculiar velocities.

While joint Likelihood analyses employing both CMB and LSS data are
allowing more accurate estimates of cosmological
parameters, they involve various subtle statistical issues:
\begin{itemize}
\item The choice of the model parameter space is somewhat arbitrary.
\item One commonly solves for the probability for the data given a model
      (e.g. using a Likelihood function),  
      while in the Bayesian framework this should be modified
      by the prior for the model.
\item If one is interested in a small set of parameters, should one marginalise
      over all the remaining parameters, rather than  fix them at certain 
      (somewhat ad-hoc) values ?  
\item The `topology' of the Likelihood contours may not be simple. 
      It is helpful when the Likelihood contours of different probes 
      `cross' each other to yield a global maximum 
       (e.g. in the case of CMB and SNe), but in other cases
       they may yield distinct separate `mountains', and the joint 
       maximum Likelihood may lie in a `valley'.
\item Different probes might be spatially correlated, i.e. 
       not necessarily independent.
\item What weight should one give to each data set ?
\end{itemize} 

The above points have been discussed in many papers in the cosmological
literature and also at this conference.
Here we focus on the last point.
A conventional approach does not take into account the fact that
different systematics may affect each data set. The problem arises when
data sets are inconsistent with one another. One approach is to combine
inconsistent data sets in the hope that the various systematic effects
will tend to cancel out.
However, this
may lead to problems if all of parameter space is ruled out by one data
set or another. The orthogonal approach is to choose, 
somewhat ad-hoc, a mutually,
consistent group of data sets to combine. 
Lahav et al. (2000; hereafter L2000)  presented a more objective method
for dealing with disagreement between data sets
by utilising `Hyper Parameters'
(hereafter HPs).
Some previous approaches to this problem 
of assigning  the relative weights of different measurements
have been suggested in the astronomical literature
(e.g. Godwin \& Lynden-Bell 1987; Press 1996).

The derivation of HPs is  given 
in Section 2. In Section 3 we apply the method to the 
recent Boomerang and Maxima CMB 
experiments, and we estimate the best fit 
Hubble constant ($H_0 = 100 h $ km/sec/Mpc).
Extensions of the methods are discussed in Section 4. 

\section{`Hyper-Parameters'}

Assume that we have two independent data sets, $D_{A}$ and $D_{B}$
(with $N_{A}$ and $N_{B}$ data points respectively) 
and that we wish to determine a vector of free parameters ${\bfw}$
(such as the density parameter $\Omega_{\rm{m}}$, the Hubble constant $H_0$ etc.).
This is commonly done by minimising  
\begin{equation}
\chi^2_{\rm joint} = \chi^2_A \; + \; \chi^2_B\; , 
\label{chi2_simple}
\end{equation}
(or, more generally,  maximizing 
the product of 
Likelihood functions).

Such procedures assume that the quoted observational random errors 
can be trusted, and that the two (or more) $\chi^2$s  
have equal weights.  
However, when combining `apples and oranges' one may wish to allow freedom in 
the relative weights. 
One possible approach is to generalise Eq. 1 to be 
\begin{equation}
\chi^2_{\rm joint} = \alpha  \chi^2_A \; + 
                \beta \; \chi^2_B \; , 
\label{chi2_hp}
\end{equation}
where $\alpha$ and $\beta$ are
 `Hyper-Parameters',
which are   to be dealt with 
the following Bayesian way.
There are a number of ways to interpret the meaning of the HPs.
One way is to understand $\alpha$ and $\beta$ as
controlling the relative weight of the two data sets.
It is not uncommon that astronomers 
accept and discard measurements (e.g. 
by assigning $\alpha=1$ and $\beta=0$)
in an ad-hoc way. 
The procedure proposed by L2000 gives an objective diagnostic 
as to which measurements are problematic 
and deserve further understanding of systematic or random 
errors.   
A simple example of the HPs is the case
that 
\begin{equation}
\chi^2_{A} = 
\sum {\frac {1} {\sigma_{i}^{2}}}  [x_{{\rm obs},i} - x_{{\rm pred},i}(\bfw)]^2\;,
\label{chi2_example}
\end{equation}
where the sum is over $N_{A}$ 
measurements and corresponding predictions and errors
$\sigma_{i}$. Hence by multiplying $\chi^2$ by $\alpha$
we may interpret each  error 
as effectively 
becoming ${\alpha}^{-1/2} \sigma_{i}$.

How do we eliminate the unknown HPs  $\alpha$ and $\beta$ ?
L2000 followed  the Bayesian formalism 
given (in other contexts) in Gull (1989), MacKay (1992), 
Bishop (1995) and Sivia (1996). 
By marginalisation over $\alpha$ and $\beta$ we can write
the probability for the parameters $\bfw$ given the data:
\begin{equation}
P(\bfw| D_{A}, D_{B}) = \int \int P(\bfw, \alpha, \beta | D_{A}, 
D_{B})  \; d \alpha \;d \beta \;.
\end{equation}
Using Bayes' theorem we can write the following relations:
\begin{equation}
P(\bfw, \alpha, \beta | D_{A}, D_{B}) = 
\frac {P(D_{A}, D_{B} | \bfw, \alpha, \beta) \;P(\bfw, \alpha, \beta) } 
{P(D_{A}, D_{B})} \;,
\end{equation} 
and 
\begin{equation}
P(\bfw, \alpha, \beta) = P(\bfw | \alpha, \beta) \; P(\alpha, \beta) \;. 
\end{equation} 
We now make the following assumptions:
\begin{equation}
P( D_{A},D_{B}|\bfw, \alpha, \beta) = 
P(D_{A}| \bfw,\alpha) \; P(D_{B}|\bfw,\beta) \;,
\end{equation} 

\begin{equation}
P(\bfw |\alpha, \beta) = {\rm const.} \;,
\end{equation}

\begin{equation}
P(\alpha, \beta) = P(\alpha) \; P(\beta) \;.
\end{equation} 
With the choice of `non-informative' uniform priors in the log, 
$P(\ln \alpha) = P(\ln \beta) =1$  
we get
$P(\alpha) = 1/\alpha$  and $P(\beta) = 1/\beta$ (Jeffreys 1939).
Note that the integral over priors of this kind
diverges
(such a prior is called `improper', see Bishop 1995).
These are very conservative priors, essentially 
stating that we are ignorant about the 
scale of measurements and  errors.
The other extreme is obviously $P(\alpha) = \delta(\alpha-1)$, 
i.e. when the measurements and  errors are taken faithfully. 
One can try other forms (see below), 
but it is likely that these two extreme forms 
reasonably bracket the probability space.  
Hence:
\begin{equation}
P(\bfw| D_{A}, D_{B}) =  { 1 \over P(D_{A}, D_{B}) } \; P(D_{A} | \bfw) 
\;P(D_{B} | \bfw), 
\label{PwDADB}
\end{equation} 
where 
\begin{equation}
P(D_{A} | \bfw) \equiv \int P(D_{A} | \bfw, \alpha) \alpha^{-1} d \alpha 
\;,
\label{PDA}
\end{equation} 
and
\begin{equation}
P(D_{B} | \bfw) \equiv \int P(D_{B} | \bfw, \beta)  \beta^{-1} d \beta \;.
\label{PDB}
\end{equation}

 It is common to have a  
 likelihood function of the form of a Gaussian in $N_{A}$ 
 dimensions:
 \begin{equation}
 P_{G}(D_{A}|\bfw) \; \propto \;\exp[ -\chi_{A}^2 /2]\;,  
\label{mvGauss}
\end{equation}
where we assume  
for simplicity that the normalization 
constant is independent  of the parameters
$\bfw$ (this is indeed the case in our application for the CMB 
measurements in the next Section).

We generalise this form to incorporate $\alpha$ as follows:
\begin{equation}
P(D_{A} | \bfw, \alpha) \; \propto \; \alpha^{N_{A}/2}\; 
\exp (-{\alpha \over 2} \chi_{A}^{2} ) \; .
\label{Gauss_alpha}
\end{equation} 
The integral of Eq. 11 
then gives 
\begin{equation}
P(D_{A} | \bfw) \propto  
{(\chi_{A})}^{-N_{A}}\;,
\label{power-law}
\end{equation}
and similarly for Eq. 12. 
We note that it is the specific choice of prior for $P(\alpha)=1/\alpha$
that has led to a change from a  Gaussian distribution 
(Eq. 13)
to a power-law 
(Eq. 15).
Eq. 10
can then be written (ignoring constants) as
\begin{equation}
-2 \; \ln P(\bfw| D_{A}, D_{B})\; = 
N_{A} \ln (\chi_{A}^{2})   
\; + \;   N_{B} \ln (\chi_{B}^{2})\;. 
\label{lnPwDADB}
\end{equation} 
To find the  best fit parameters $\bfw$ requires us to minimise
the above probability in the $\bfw$ space.
Note that in this case our method is equivalent to assuming that we are
ignorant of the relative scale of the errors in each experiment.
It is as easy to calculate this statistic as the standard $\chi^2$.
Eq. 16 
actually 
generalises a similar
equation derived by Cash (1979) using an entirely different set of assumptions.

Since $\alpha$ and $\beta$
have been eliminated from the analysis by
marginalisation they do not have particular values that can be quoted.
Rather, each value of $\alpha$ and $\beta$ has been considered and weighted
according to the probability of the data given the model.
However, it may be useful to know which
values of $\alpha$ and $\beta$
were given the most weight. This can be
estimated by finding the values of 
$\alpha$ and $\beta$ at which 
Eq. 14
peaks:
\begin{equation}
\alpha_{\rm {eff}} = \frac { N_{A}} {\chi_{A}^{2}}\;,  
\label{alpha1}
\end{equation} 
and similarly
\begin{equation}
\beta_{\rm {eff}} = \frac { N_{B}} {\chi_{B}^{2}} \;, 
\label{beta1}
\end{equation} 
both evaluated at the joint peak.
We note that if we substitute these effective $\alpha$ and $\beta$
in Eq. 2 
we obtain $\chi^{2}_{\rm joint} = N_{A} + N_{B}$. 

There is of course freedom in choosing the prior. 
For example, if we take $P(\alpha) =1 $ (instead of Jeffreys' prior
$P(\alpha) = 1/\alpha$) we find that the function to be minimised is
\begin{equation}
-2 \; \ln P(\bfw| D_{A}, D_{B})\; = 
(N_{A}+2) \ln (\chi_{A}^{2})   
\; + \;   (N_{B}+2) \ln (\chi_{B}^{2})\; 
\label{lnPwDADB}
\end{equation} 
instead of Eq. 16.
Thus these two priors give very similar results for large $N_A$.
Numerous other priors are possible (e.g. a top-hat centred on 
a plausible value), but at the expense of more free HPs
(e.g. the width of the top-hat).
Illustrations of the HPs approach applied to 
toy-models are given in Bridle (2000), and another application 
of the above HPs  
(to galaxy cluster data)
is given in Diego et al. (2000).


\section {Application to the CMB Data}

\subsection {The Boomerang and Maxima Data}

The recent Boomerang (hereafter B; de Bernardis et al. 2000) 
and Maxima (hereafter M; Hanany et al. 2000) CMB anisotropy measurements 
yielded  high-quality angular power spectra $C_l$ over the 
spherical harmonics $ 400 \simlt  l \simlt  800$.
An important factor in interpreting the data is the calibration error.
The experimental papers quote calibration errors of 10\% and 4\%
(1-sigma in $\Delta T/T$) for B and M, respectively.
The measurements (with B data corrected upward by 10\%, 
and M data corrected downward by 4 \%) 
are shown in Figure 1, and they indicate a  
well defined  first acoustic peak at $l \sim 200$, with less convincing 
second and third peaks at higher harmonics.
These measurements favour (under certain assumptions) 
a flat universe, spectral index $n=1$ 
and baryon density $\Omega_b h^2 \sim 0.03$ 
(e.g. Jaffe et al. 2000; Bond et al. 2000; Bridle, this volume), 
which is about 2-sigma
higher than the Big-Bang Nucleosynthesis (BBN) value 
$\Omega_b h^2 \sim 0.0190 \pm 0.0018$ (95 \% CL; Burles et al. 2000).
Note that the recent CBI result (Padin et al. 2000) gives a higher
power (at $l\sim 600$) relative to  B\&M.
Jaffe et al. (2000) fitted models after 
combining the B\& M data sets into one set. 
Here we take a different approach for joint analysis of the two data sets 
by utilising the `Hyper-Parameters'.

\begin{figure}  
\plotone{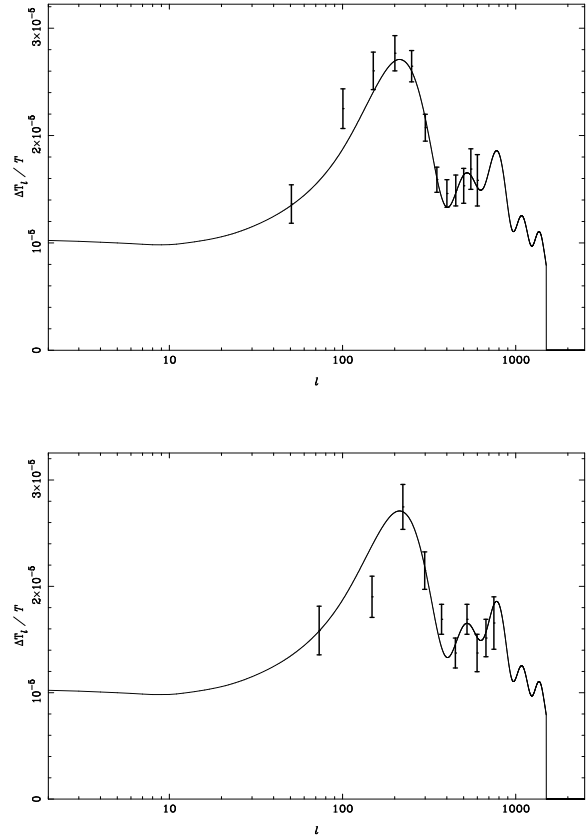}
\caption{The Boomerang $ \Delta T \over T$ data (top panel,
calibrated by 1.10) and Maxima data (bottom panel, calibrated by
0.96). The line in each panel is for a $\Lambda$-CDM model with $n=1,
\Omega_m = 1 -\Lambda = 0.3, \Omega_b h^2 = 0.03, \qrms = 18 \mu K$,
and our `best fit' $h = 0.8$.}
\end{figure}

\subsection {Results}

We illustrate the  effect of using HPs by application to
measurements of the angular power spectrum of the Cosmic Microwave
Background (CMB). 
Numerous groups have now used CMB data to
estimate cosmological parameters. 
The most common method is the flat bandpower method (Bond
1995) in which the difference between observed and
predicted flat bandpowers are compared using the
$\chi^2$ statistic (Eq. 3). 
We note that non-zero correlations between the CMB data points can make
the data points look more smooth which, since the theoretical model is
smooth on this scale, will tend to improve the apparent goodness of fit
to the model and thus inappropriately give more weight to correlated data
points. We also note that the assumption that the 
Likelihood function is a Gaussian is only an approximation 
(Douspis et al. 2000).  


L2000 applied the HPs approach to the pre-B\&M 
CMB data sets, in different combinations. 
Here we apply the method  to the recent high-quality B\&M data, 
first in their `raw' form and then in their calibrated form.
For simplicity, we restrict ourselves to a very limited
set of cosmological models. 
We obtain theoretical CMB power-spectra using the CMBFAST and 
CAMB codes (Slejak \& Zaldarriaga 1996; Lewis, Challinor \& Lasenby 2000).
We assume that CMB fluctuations arise from
adiabatic initial conditions with Cold Dark Matter (CDM) and negligible
tensor component, in a flat Universe with  $\Omega_{\rm{m}}=0.3$,
$\lambda=1-\Omega_{\rm{m}}=0.7$, $n=1$,
$\qrms=18\mu$K and $\Omega_b h^2 = 0.03$.
This choice is motivated by numerous other studies which combined
CMB data with other cosmological probes (e.g. Jaffe et al. 2000,
Bridle et al. 2000; Hu et al. 2000).  
We then
investigate the constraints on the remaining parameter, the
dimensionless Hubble constant, $h=H_0/(100$~\mbox{$\rm{km} \rm{s}^{-1}
\rm{Mpc}^{-1}$}$)$. 
Increasing
$h$ decreases the height of the first acoustic peak, and makes few
other significant changes to the angular power spectrum (e.g. Hu et al. 2000). 
The range
in $h$ investigated here is ($0.5<h<1.1$).
The results using conventional $\chi^2$ 
(Eqs. 1 and 3) 
are shown in Table 1, and with the HPs approach (Eq. 16) in Table 2.
The full likelihood functions are given in Figures 2 and 3.
We see that the raw (uncalibrated) B\&M data give two distinct values
in the standard $\chi^2$ analysis. The HPs approach on the raw data suggests
that B carries 4.5 times more weight than M (the ratio of the HPs), 
for this particular choice of model and parameter space, yielding
a best $h=0.88$.
However, the calibration of the data (as described in the caption to Table 1) 
brings the two data
sets to much better agreement (e.g. the ratio of the B/M HPs is now 1.3).
In fact, in this case the standard joint $\chi^2$ and the HPs (for two different choices of priors; Eqs. 16 and 19)
give the same result, $h=0.79$, with slightly 
smaller error bars in the HPs case ($\pm 0.04;  95\%$ CL).
This best fit model is shown in Figure 1.   
We also tried the BBN value 
$\Omega_b h^2 = 0.019$ 
(last entries in Table 1 and 2), which we can see
gives much poorer $\chi^2$ than the value
$\Omega_b h^2 = 0.03$  
(as also suggested 
by Jaffe et al. 2000 and others).

\begin{figure}  
\plotone{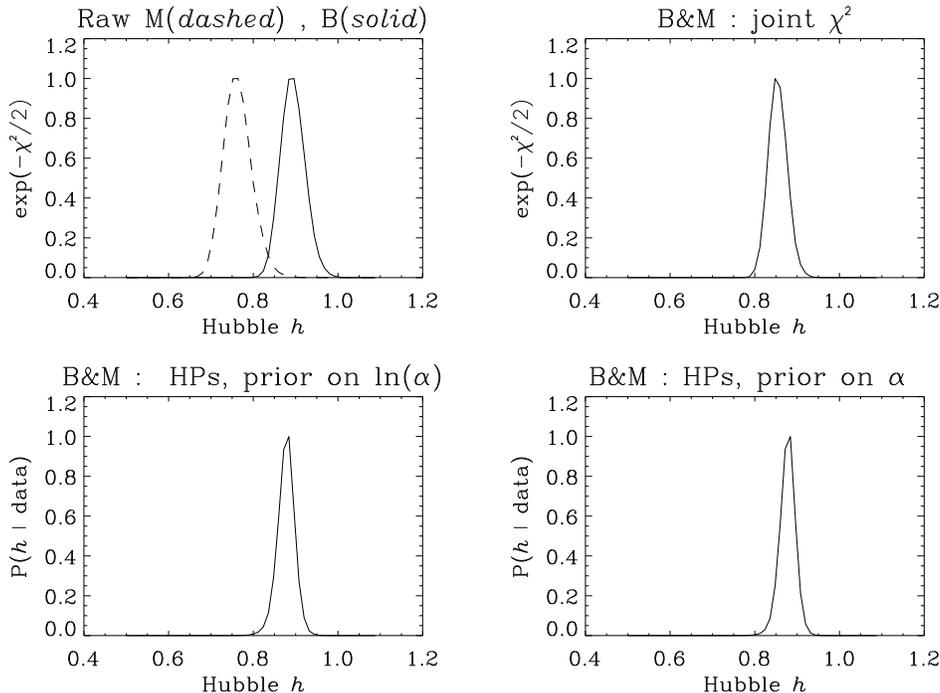}
\caption{Analysis with raw (uncalibrated) Boomerang \& Maxima
data.  The top-left
panel is for the individual $\chi^2$ of each data sets (eq. 3), while
the top-right panel is for the sum of $\chi^2$ (Eq. 1).  The
bottom-left panel is the Hyper-Parameters probability with the prior
$P(\ln \alpha) = 1$ (Eq. 16), and the bottom left panel is for the
prior $P(\alpha) =1$ (Eq. 19).}
\end{figure}

\begin{figure}  
\plotone{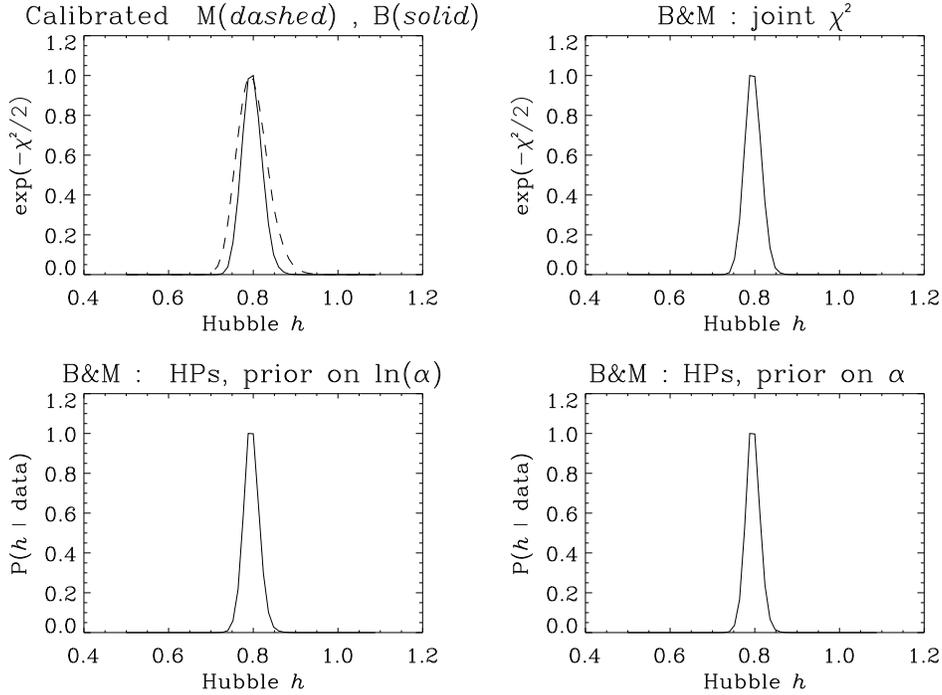}
\caption{Analysis with calibrated B \& M data 
(by 1.10 and 0.96 in $\Delta T \over  T$, respectively).
The panels are as in Figure 2. Note the good agreement between the two data 
sets, and between the joint $\chi^2$ approach and the HPs approach 
(for the two different priors).}
\end{figure}

\begin{table}
\center{\vbox{
\begin{tabular}{@{}lrrr}
\hline
Data      &        $N$ & Best $h$ &   $\chi^2$ \\
\hline
BOOMERANG (raw) & 12     & $0.90$ & 4.4 \\
\hline
MAXIMA (raw)   & 10     & $0.76$ & 7.1 \\
\hline
B\&M (raw)      & 22     & $0.85$ & 19.8 \\
\hline
BOOMERANG (calibrated by 1.10)& 12     & $0.80$ & 8.8 \\
\hline
MAXIMA  (calibrated by 0.96)  & 10     & $0.79$ & 8.9 \\
\hline
B\&M  (calibrated)  & 22    & $0.79$ & 17.7 \\
\hline
B\&M  (calibrated, BBN)  & 22    & $0.72$ & 35.7 \\
\hline
\end{tabular}
}}
\label{tableeachalone}
\caption{
Conventional $\chi^2$ analysis using each B and M data set alone, 
and both sets combined. 
Results are given for `raw' (uncalibrated) data, and for calibration 
of $\Delta T /T$ by factors of 1.10 and 0.96 for B \& M respectively,
as explained in the text.
For each data set the number of data points, $N$, the best fit
value of $h$ and the $\chi^2$ value at this point are given. The full
likelihood distributions in $h$ are shown in Figures 2 and 3.
Other parameters are fixed for a $\Lambda$-CDM model at $\Omega_{\rm{m}}=0.3$,
$\lambda=1-\Omega_{\rm{m}}=0.7$, 
$n=1$,
$\qrms=18\mu$K and $\Omega_b h^2 = 0.03$.
For the last entry  $\Omega_b h^2 = 0.019$ (BBN value).
}
\end{table}
\begin{table}
\center{\vbox{
\begin{tabular}{@{}lrrr@{}l@{}lr@{}l@{}l}
\hline
 Data & $N$ & Best $h$ &&& Effective HP \\
\hline
B\&M (raw)    & 22  & 0.88    &&&   2.7(B); 0.6 (M) \\
\hline
B\&M (calibrated)  & 22  & 0.79  &&&  1.4(B); 1.1 (M)  \\
\hline
B\&M (calibrated, BBN)  & 22  & 0.73  &&&  0.5(B); 0.7 (M)  \\
\hline
\end{tabular}
}}
\label{table}
\caption{
The results of the Hyper-Parameters
analysis (Eq. 16) The data sets are as described in Table 1.
Shown are 
the number of data points $N$ in each data set, 
the best fitting value of $h$, 
and the effective HP ($N/\chi^2$)
at this $h$. Other parameters were held fixed as 
described in Table 1.
}
\end{table}

\section {Discussion}

We have presented a formalism for analysing a combination of
measurements, when it is likely that different systematics (or
methods for calculating random errors)
may affect each data set differently.
By using a Bayesian analysis, and by using 
a specific 
`non-informative' prior
for the `Hyper-Parameters' ($P(\ln \alpha)=1$), 
we find that for $M$ data sets  one 
should minimise 
\begin{equation}
-2 \; \ln P(\bfw| {\rm data} ) = 
\sum_{j=1}^{M} N_{j} \ln (\chi_{j}^{2}),   
\label{sum_hp}
\end{equation}  
where $N_j$ is the number of measurements in data set $j=1, ..., M$.
It is as easy to calculate this statistic as the standard $\chi^2$.
The corresponding HPs $\alpha_{{\rm eff},j} = N_j/\chi^2_j$ 
provide useful diagnostics on the reliability of different data sets. 
We emphasize that a low HP assigned to an experiment does not necessarily
mean that the experiment is `bad', but rather it calls attention to look 
for systematic effects or better modelling. 

In L2000 we analysed pre-B\&M data and found that 
while the standard $\chi^2$ approach 
gave a wide range for $H_0$, the Hyper-Parameter analysis
suggested two distinct values of $H_0$, $\sim 50$ and $\sim 70$
km/sec/Mpc. 
Here we applied the method to the B \& M data, with and without calibration.
The HPs indeed `detect' inconsistencies between the two `raw' data sets,
but the calibrated data sets show good agreement with each other, as seen 
in both the $\chi^2$ and the HPs statistics. We have also seen in this example 
that the HPs solution is insensitive to the exact choice of prior.  

The best fit Hubble constant is $H_0 = 79 \pm 4$ km/sec/Mpc 
(95\% CL, random errors only)
for a fixed flat CDM $\Omega_{\rm m} = 1 - \lambda = 0.3$ model with
$n=1$,
$\qrms=18\mu$K and $\Omega_b h^2 = 0.03$.
We note that if more cosmological parameters are left free and then 
marginalised over, the error in $h$ would typically be larger
(e.g. Bond, Bridle in this volume). 

This combination of $\Omega_m$ and $H_0$ corresponds gives for 
the age of the Universe  11.9 Gyr.
Our derived $H_0$ is slightly higher 
but still
 consistent with the `final result' of $H_0$ from Cepheids
and other distance indicators 
(Freedman et al. 2000)
$H_0  = 72 \pm (3)_r \pm (7)_s$ km/sec/Mpc 
(1-sigma random and systematic errors).

The above analysis can be extended in a number of ways.
Current and future CMB data  can be 
combined with other cosmological probes
(and their corresponding HPs), and more cosmological 
parameters can be kept free.
Here we used a simple correction for the calibration error.
A more general approach is to marginalise over both the HPs and a 
calibration probability function (Bridle et al, in preparation).
Two other aspects which can be modified
according to specific problems are the priors $P(\alpha_j)$ and
the probability functions $P(D_j|\bfw)$. We shall discuss
these extensions elsewhere.

{\bf Acknowledgments.} 
I thank my collaborators S. Bridle, M. Hobson, A. Lasenby and
L. Sodre for their contribution, and to L. Page and J. Mould for 
helpful discussions.








\end{document}